\documentclass{iaus}
\usepackage{graphicx}

\title[Diagnostics for spectropolarimetry and magnetography] 
{Diagnostics for spectropolarimetry and magnetography}

\author[Del Toro Iniesta \& Mart\'{\i}nez Pillet]   
{Jose Carlos del Toro Iniesta$^1$
 \and Valent\'{\i}n Mart\'{\i}nez Pillet$^2$}

\affiliation{$^1$Instituto de Astrof\'{\i}sica de Andaluc\'{\i}a (CSIC), \\ Apdo. de Correos 3004,
E-18080, Granada, Spain \\ email: {\tt jti@iaa.es} \\[\affilskip]
$^2$Instituto del Astrof\'{\i}sica de Canarias, \\ V\'{\i}a L\'actea, s/n, E-38200, La Laguna, Spain \\email: {\tt vmp@iac.es}}

\pubyear{2011}
\volume{273}  
\pagerange{100--126}
\jname{Physics of Sun and Star Spots}
\editors{D.P. Choudhary \& A.C. Cadavid, eds.}
\begin{document}

\maketitle

\begin{abstract}
An assessment on the capabilities of modern spectropolarimeters and magnetographs is in order since most of our astrophysical results rely upon the accuracy of the instrumentation and on the sensitivity of the observables to variations of the sought physical parameters. A contribution to such an assessment will be presented in this talk where emphasis will be made on the use of the so-called response functions to gauge the probing capabilities of spectral lines and on an analytical approach to estimate the uncertainties in the results in terms of instrumental effects. The Imaging Magnetograph eXperiment (IMaX) and the Polarimetric and Helioseismic Imager (PHI) will be used as study cases.
\end{abstract}

\firstsection 
\section{Introduction}

Modern solar spectropolarimeters and magnetographs are vectorial because all four Stokes parameters of the light spectrum are measured. Longitudinal magnetography (i.e., Stokes $I \pm V$) can be interesting for some specific applications, but the partial analysis is usually included (if possible) as a particular case of the more general, full-Stokes polarimetry. Some of these modern instruments have been recently or are currently in operation (e.g., the Tenerife Infrared Polarimeter, TIP, Mart\'{\i}nez Pillet et al. 1999, Collados et al. (2007); the Diffraction-Limited Spectro-Polarimeter, DLSP, Sankarasubramanian et al. 2004; CRISP, Narayan et al. 2008; the Imaging Magnetograph eXperiment, IMaX, Mart\'{\i}nez Pillet et al. 2010, for the {\em Sunrise} mission, Bartol et al. 2010; and the Helioseismic and Magnetic Imager, HMI, Graham et al. 2003, for the {\em Solar Dynamics Observatory} mission, Title 2000), some other are being designed and built for near future operation and missions (e.g., the Polarimetric and Helioseismic Imager, SO/PHI, [formerly called VIM, Mart\'{\i}nez Pillet, 2006] for the {\em Solar Orbiter} mission, Marsch et al. 2005). Assessing their capabilities in terms of their accuracy for retrieving the solar line-of-sight (LOS) velocity ($v_{\rm LOS}$) and vector magnetic field (of components $B$, $\gamma$, and $\phi$) is in order since such an analysis can diagnose how far reaching is our current and near-future understanding of the solar atmosphere. The diagnostics is relevant both for the design of new instruments in order to maximize their performances and for the analysis of uncertainties in data coming from currently operating devices. Certainly, no fully general assessment can be devised that includes all possible polarimeters and a family of them should be considered in each specific study. Here we restrict our analysis to those spectropolarimeters and magnetographs whose polarization modulator consists of two nematic liquid crystal variable retarders (LCVRs). Hopefully, the discussion presented in this invited contribution helps further diagnostics of other instruments.

\section{Rules for improving the measurements}
\label{sec:rules}

Since we only measure photons, every inference we can make out of the observations naturally depends on photometric accuracy. Assuming that systematic errors are under control (ideally absent) two are, therefore, the pillars which the quality of measurements rests upon: the signal-to-noise ratio ($S/N$) and the minimum variations, $\delta S_i$,  that the Stokes parameters exhibit after a perturbation in the solar physical quantities.\footnote{We shall hereafter denote by ${\bf S} = (S_1, S_2, S_3, S_4)$ the vector of Stokes parameters, usually called $(I, Q, U, V)$.} If the latter are larger than the uncertainties in the Stokes signals due to noise, the measurements are useful. Otherwise, they are not. One should, then, design new instruments so that $S/N$ and $\delta S_i$ are maximized and results from current instruments are more accurate wherever these quantities are larger. 

\subsection{Increasing the signal-to-noise ratio}
\label{sec:increasingston}

As a first tool for improving $S/N$, modern polarimeters introduce image accumulation of $N_a$ individual exposures. Besides, every Stokes parameter is obtained from $N_p$ polarization modulation states, so that a total of $N_p N_a$ individual frames contribute to a given Stokes parameter image. If $\overline{\sigma}_i$ stands for the individual frame contribution to $\sigma_i$, the uncertainty in $S_i$ then is
\begin{equation}
\label{eq:errorsigmai}
\sigma_i = \overline{\sigma}_i \, \sqrt{N_p N_a},
\end{equation}
where we have assumed photon noise. 

According to Mart\'\i nez Pillet et al. (1999) and to Del Toro Iniesta \& Collados (2000), 
\begin{equation}
\label{eq:errorefficiency}
\overline{\sigma}_i = \frac{\sigma}{\epsilon_i},
\end{equation}
where $\sigma$ is the noise-induced uncertainty for each individual exposure and $\epsilon_i$ is the so-called polarimetric efficiency for Stokes $S_i$. Then, it is easy to see that, if $s/n$ denotes the signal-to-noise ratio of each individual exposure, 
\begin{equation}
\label{eq:ston1}
\left(S/N\right)_i = (s/n)\, \epsilon_i \, \sqrt{N_pN_a}.
\end{equation}

Equation (\ref{eq:ston1}) tells us that the larger the polarimetric efficiencies and/or the larger the number of individual exposures, the larger the signal-to-noise ratio for each Stokes parameter. $N_p$ is often (advisably) kept to its minimum value of 4 in order to preserve integrity in the Stokes analysis in a minimum time. This can only be done, however, when the polarization modulator permits it as, indeed, in our LCVR-based polarimeters, but it seldom exceeds 6 or 8. The number of accumulations is usually traded-off with the solar dynamic time scales, in order not to blur information on time-evolving solar features with a too long effective exposure time. Then, optimization of polarimetric measurements basically lies in maximization of polarimetric efficiencies. According to Del Toro Iniesta \& Collados (2000), an ideal polarimeter wishing to have equal signal-to-noise ratios for Stokes $S_2$, $S_3$, and $S_4$ can reach maximum efficiencies given by
\begin{equation}
\label{eq:maxefficiencies}
\epsilon_1 = 1, \,\, \epsilon_{2,3,4} = 1/\sqrt{3}.
\end{equation}

Since we usually speak of only one (``the") signal-to-noise ratio of the observations, we are implicitly meaning $(S/N)_1$, that is, the signal-to-noise ratio for the intensity. Therefore, after Eq. (\ref{eq:ston1}), we can write
\begin{equation}
\label{eq:ston2}
\left(S/N\right)_i = \frac{\epsilon_i}{\epsilon_1} \left(S/N\right),
\end{equation}
so that, if we have for instance $S/N = 10^3$, then $(S/N)_{2,3,4} \leq 577$, according to Eqs. (\ref{eq:maxefficiencies}). It is important to point out that we are speaking about single wavelength samples. Every observational quantity involving several samples can certainly improve $(S/N)_{2,3,4}$ above this limit. Although simple, the result in Eq. (\ref{eq:ston2}) has not ever been brought to the attention of the community as far as we know, and is paramount to assessing observational accuracies: polarimetry imposes an extra penalty in terms of $S/N$ as compared to normal spectroscopy or photometry. Such a penalty roots in the differential character of polarimetric measurements. A discussion on how optimum polarimetric efficiencies can be reached (at least theoretically) with LCVR-based polarimeters is deferred to Sec. \ref{sec:efficiencies}.

\subsection{Maximizing the spectral line sensitivities}
\label{sec:sensitivities}

As explained in the beginning of Sect. \ref{sec:rules}, the other ingredient for improving measurements quality is the spectral line sensitivity. Increasing $\delta S_i$ can only be achieved by carefully selecting the line. Fortunately, the tools for such a selection are at our disposal as well. As explained by Ruiz Cobo \& Del Toro Iniesta (1994; see references to pioneering work over there), the sensitivity of Stokes profiles to perturbations in the solar physical quantities are directly given by the response functions (RFs). We insist on the importance of perturbations: we can only discern different LOS velocities or field strengths in two structures provided the modification in the Stokes profiles are large enough (that is, larger than the threshold imposed by noise) in one of the structures as compared to the other. It is perturbation theory the technique that enables us to evaluate how large $\delta S_i$ are for given variations in $v_{\rm LOS}$, $B$, $\gamma$, and/or $\phi$. RFs are defined such that, for every single wavelength, 
\begin{equation}
\label{eq:rfdefinition}
\delta S_{i_x} = \int_{-\infty}^{+\infty} R_{i_x} \, \delta x \, \mbox{d} \tau,
\end{equation}
where $x$ is an index representing either one of the physical quantities of interest, $\tau$ stands for the optical depth, and $R_{i_x}$ is the response function of $S_i$ to perturbations in $x$. Naturally, and within a linear approximation, $\delta S_i = \sum \delta S_{i_x}$, the sum being extended to all the quantities. Equation (\ref{eq:rfdefinition}) paves the way for estimating detection thresholds for the different physical quantities. For example, the detectable two-sigma field strength, $\delta B_{\rm min}$, would be such that $\delta S_{i_{B_{\rm min}}} = 2 \sigma_i$. 

Analytic expressions for RFs are available under the Milne-Eddington (ME) approximation (Orozco Su\'arez \& Del Toro Iniesta, 2007) that are useful even for more accurate estimates of uncertainty levels (Del Toro Iniesta, Orozco Su\'arez, \& Bellot Rubio, 2010) accounting for details of the specific technique used to retrieve given quantities. But purely phenomenological approaches are also valid to establish real rankings of spectral lines according to their ability for inferring velocities, magnetic fields, and so on (Cabre\-ra Solana et al., 2005). Finally, a further approach to determine which particular line is more useful for being used with a given instrument has recently been provided by simulations. MHD simulations are a modern and useful tool to elaborate educated guesses of instrument behavior since real observations can be computationally reproduced. This has been the way, for instance, how Orozco Su\'arez et al. (2010) have been able to gather evidence in favor of the Fe {\sc i} line at 525.02 nm against that at 525.06 nm, which was originally foreseen for the IMaX instrument (see details in Mart\'{\i}nez Pillet et al. 2010).

\subsection{Detection thresholds}
\label{sec:thresholds}

Scientific requirements on given physical quantities can be translated into instrument requirements, provided an inference technique to retrieve that quantity is known. This is an advisable exercise that helps for trading-off the many instrumental parameters that must be taken into account during the design phase. As a first example, imagine we are going to infer LOS velocities through a Fabry-P\'erot spectrometer and the Fourier tachometer technique. Such a technique involves four Stokes $S_1$ samples that combined, according to Fernandes (1992), give
\begin{equation}
\label{eq:fourier}
v_{\rm LOS} = \frac{2c\,\delta \lambda}{\pi\lambda_0} \arctan \frac{I_{-9} + I_{-3} - I_{+3} - I_{+9}}{I_{-9} - I_{-3} - I_{+3} + I_{+9}},
\end{equation}
where $c$ stands for the speed of light, $\delta \lambda$ is the \'etalon spectral resolution, $\lambda_0$ is the central wavelength of the line, and $I_{\pm i}$ represent the Stokes $S_1$ (Stokes $I$) samples at the given wavelengths in picometers.

Error propagation in Eq. (\ref{eq:fourier}) can be shown to give the LOS-velocity expected uncertainties in terms of the \'etalon roughness, $\sigma_{\delta\lambda}$, of the thermal and voltage instabilities of the spectrometer, $\sigma_T$ and $\sigma_V$, and on the noise of the observations, $\sigma_1$. Without entering into details of the (easy but) lengthy calculations, the variance of the retrieved velocities can be written as
\begin{equation}
\label{eq:errorvelocities}
\sigma_{v_{\rm LOS}}^2 = f(v_{\rm LOS}, \delta\lambda) \sigma_{\delta\lambda}^2 + g(v_{\rm LOS}, \lambda_0, \delta\lambda, I_i, s_i) (k_T^2 \sigma_T^2 + k_V^2 \sigma_V^2) + h(\lambda_0, \delta\lambda, I_i) \sigma_1^2,
\end{equation}
where $f$, $g$, and $h$ are given functions of the specified variables, $s_i$ represent the Stokes $S_1$ profile derivatives with respect to wavelength at the sample wavelengths, and $k_T$ and $k_V$ are the temperature and voltage calibration constants for tuning the \'etalon, respectively. Assume now, for example, that $\lambda_0 = 617.3$ nm and $\delta\lambda = 100$ m\AA\ as for the SO/PHI instrument. Then, an \'etalon roughness leading to $\sigma_{\delta\lambda} = 1$ m\AA\ produces $\sigma_{v_{\rm LOS}} = 1$ m$\,$s$^{-1}$ for velocities of 100 ms$^{-1}$ (and is linear in $v_{\rm LOS}$); pure photon noise with $S/N = 10^3$ induces $\sigma_{v_{\rm LOS}} = 7$ m$\,$s$^{-1}$ or, in other words, a scientific requirement on $v_{\rm LOS}$ stability of 1 m$\,$s$^{-1}$ (that can be of interest for low-$l$, global helioseismology) demands a stability of 0.55 mK in temperature or 42 mV in voltage! Only state-of-the-art technology can aim at such thermal stabilities in a space environment, but the voltage requirement is very stringent as well since LiNbO$_3$ \'etalons are tuned with voltages of the order of $10^3$ V.

Take now as a second example the inference of longitudinal and transverse field strengths through the magnetograph equations
\begin{equation}
\label{eq:magnetograph}
B_{\rm lon} \equiv k_{\rm lon} \frac{V_{\rm s}}{I_{\rm c}} \,\, \mbox{and} \,\, B_{\rm tran} \equiv k_{\rm tran} \sqrt{\frac{L_{\rm s}}{I_{\rm c}}},
\end{equation}
where $k_{\rm lon}$ and $k_{\rm tran}$ are calibration constants, $V_{\rm s}$ and $L_{\rm s}$ are the circular and linear polarization magnetographic signals,
\begin{equation}
\label{eq:magsignals}
V_{\rm s} \equiv \frac{1}{n_\lambda} \sum_{j=1}^{n_\lambda} |S_{4,j}|, \,\, L_{\rm s} \equiv \sum_{j=1}^{n_\lambda} \sqrt{S_{2,j}^2 + S_{3,j}^2},
\end{equation}
with $n_\lambda$ being the number of wavelength samples within the spectral line. Since $n_\lambda = 4$ for the IMaX instrument, the uncertainties in $V_{\rm s}$ and $L_{\rm s}$ are necessarily a factor 2 smaller than $\sigma_{2,3,4}$ because information from four independent wavelengths is averaged for building the magnetograms. In such a case, one can estimate photon-noise-induced uncertainties of $\sigma_{B_{\rm lon}} = 4.8$ G and $\sigma_{B_{\rm lon}} = 80$ G.

\section{Maximizing the polarimetric efficiencies}
\label{sec:efficiencies}

Once we know the effect of noise in the final inferences made with the instrument and how to improve $S/N$, that in the end turns out to maximizing efficiencies, let us check whether or not real polarimeters can (theoretically) achieve or (practically) approach the optimum polarimetric efficiencies of ideal instruments.

A rule of thumb in polarimetry is to put the polarization modulator as early in the optical path as possible so as to minimize the influence of the remaining optics: after the modulator, light is encoded and, no matter the path, can finally be analyzed properly before reaching the detector. This property has not been demonstrated, however, for polarimetric efficiencies so far. In other words, can polarimeters preserve the polarimetric efficiencies regardless of the retardations and changes of phase induced by the optics between the modulator and the analyzer? This is not a trivial question because intermediate optics might change the polarimeter's Mueller matrix in a way that would modify the efficiencies; indeed, not all polarimeters can reach the optimum efficiencies. Mart\'{\i}nez Pillet et al. (2004) pointed out that nematic-LCVR-based polarimeters can theoretically achieve optimum efficiencies for both vector and longitudinal magnetography. This fact is easy to understand as we are going to demonstrate.

Assume we have two nematic LCVRs of retardances $\rho$ and $\tau$, respectively for the first and the second one to be reached by light. Such retardances can be changed at will by simply modifying the tuning voltage of the devices. If the optical axis of the first LCVR is put at 0$^{\circ}$ with respect to the positive $S_2$ direction (X axis, for instance), the second LCVR has its axis at $\pi/4$, and the linear analyzer is at 0$^{\circ}$, then the rows of the modulation matrix (Del Toro Iniesta \& Collados, 2000) are
\begin{equation}
\label{eq:modmatrix1}
O_{ij} = (1, \cos \tau_i, \sin \rho_i \sin \tau_i, -\cos\rho_i \sin\tau_i),
\end{equation}
where index $i=1,2,3,4$ corresponds to each of the four measurements. A polarimeter having these four elements equal to the efficiencies in Eqs. (\ref{eq:maxefficiencies}) is an optimum one. Since all four $O_{i1} = 1$, Stokes $S_1$ can reach its maximum efficiency. At least four different solutions can also be found to equations resulting from making the other three components equal to $1/\sqrt{3}$. Therefore, the remaining Stokes parameters can also reach their maximum efficiencies. It is also easy to understand that the best longitudinal analysis $(S_1 \mp S_4)$ can be obtained by tuning the retardance of the first LCVR to 0$^{\circ}$ and that of the second to $\pm \pi/2$.

Let us see now what is the effect of an \'etalon in between the modulator and the analyzer as in the IMaX or SO/PHI instruments. The most general way of modeling the polarization properties of such a device is by assuming it behaves like a retarder oriented at an angle $\theta$ and with a retardance $\delta$. Now the Mueller matrix of the polarimeter gets modified because the Mueller matrix of the \'etalon, ${\bf M}_3$, has to be inserted between those of the LCVRs, ${\bf M}_1$ and ${\bf M}_2$,  and that of the analyzer, ${\bf M}_4$. The Mueller matrix of the system is then ${\bf M} = {\bf M}_4 {\bf M}_3 {\bf M}_2 {\bf M}_1$ and the modulation matrix turns modified to $O_{ij} = {\bf M}_{1j} (\tau_i, \rho_i)$. Since $M_{11} = 1$, $O_{i1} = 1,\, \forall i$. If we proceed now as before by equating $O_{i2,3,4} = 1/\sqrt{3}$, we obtain trascendental equations. Fortunately, they can be shown to have a solution and, therefore, optimum efficiencies can also be reached theoretically with these {\em real} instruments.

Besides the \'etalon, several mirrors may be needed in the design in between the modulator and the analyzer. Then, a Mueller matrix representing all the mirrors has to be inserted between ${\bf M}_3$ and ${\bf M}_4$. The effect of such an insertion can be demonstrated to be a global multiplication of all the modulation matrix elements by a constant factor. Therefore, the solutions for the transcendental equations are the same and, again, optimum efficiencies can theoretically be reached. Of course, real instruments may have modulation matrices that slightly differ from the optimum ones and calibration is always necessary.

\section{Summary and conclusions}
\label{sec:conclu}

The accuracy in line-of-sight velocity and vector magnetic fields inferred from observations roots in photometric accuracy and, hence, in $S/N$. The instruments have, therefore, to be designed so that they collect, with the best polarimetric efficiencies, as many photons as possible in wavelengths of spectral lines that are as much sensitive as possible to these physical quantities.

In this contribution we have gathered rules for increasing the $S/N$, for finding out the more sensitive spectral lines to given quantities by means of the response functions, and for deducing detectability thresholds imposed by noise. The optimization of the signal-to-noise ratio of the observations goes necessarily through the maximization of polarimetric efficiencies and we have also shown that the optimum theoretical efficiencies can be reached with nematic-LCVR-based spectropolarimeters and magnetographs like IMaX and SO/PHI.

\section*{Acknowledgements}

This work has been partially funded by the Spanish MICINN, through project AYA2009-14105-C06, and Junta de Andaluc\'{\i}a, through project P07-TEP-02687.

\begin{discussion}

\discuss{}{}

\end{discussion}

\end{document}